\documentclass[12pt,journal,a4paper,onecolumn,draftclsnofoot]{IEEEtran}
\usepackage{amssymb, amsmath, amsthm, mathtools}
\usepackage{cite}
\usepackage{pgfplots}
\usepackage{pgfplotstable}
\usepackage[caption=false,font=footnotesize]{subfig}
\usepgfplotslibrary{groupplots}
\pgfplotsset{compat=1.11}
\usepackage{amsmath,amssymb}
\usepackage{mathtools} 
\usepackage{xspace}
\newcommand{\Tableref}[1]{Table~\ref{#1}}

\newcommand{\Eqref}[1]{(\ref{#1})}

\newcommand{\Secref}[1]{Section~\ref{#1}}
\newcommand{\Figref}[1]{Fig.~\ref{#1}}

\newcommand{\ordo}{\mathcal{O}}
\newcommand{\E}{\mathbb{E}}
\newcommand{\Exp}[1]{\mathbb{E}\left[#1\right]}

\newcommand{\imag}[1]{\Im\left\{#1\right\}}
\newcommand{\real}[1]{\Re\left\{#1\right\}}

\newcommand{\trace}{\operatorname{tr}}

\newcommand{\herm}{^\text{H}}

\newcommand{\trans}{^\text{T}}
\newcommand{\inv}{^{-1}}
\newcommand{\conj}{^{*}}
\newcommand{\bx}{\mathbf{x}}

\newcommand{\bX}{\mathbf{X}}

\newcommand{\bh}{\mathbf{h}}

\newcommand{\bg}{\mathbf{g}}

\newcommand{\bw}{\mathbf{w}}

\newcommand{\by}{\mathbf{y}}
\newcommand{\bI}{\mathbf{I}}
\newcommand{\bQ}{\mathbf{Q}}
\newcommand{\bR}{\mathbf{R}}
\newcommand{\bA}{\mathbf{A}}
\newcommand{\bB}{\mathbf{B}}
\newcommand{\bC}{\mathbf{C}}

\newcommand{\be}{\mathbf{e}}

\newcommand{\bv}{\mathbf{v}}
\newcommand{\bU}{\mathbf{U}}

\newcommand{\bPhi}{\mathbf{\Phi}}

\newcommand{\bzero}{\boldsymbol{0}}
\newcommand{\mC}{\mathbb{C}}

\newcommand{\CN}{\mathcal{CN}}
\newcommand*{\iu}{\mathsf{i}}%

\DeclarePairedDelimiter\norm{\lVert}{\rVert}

\def\given{\middle|}

\newcommand{\mK}{\mathcal{K}}

\newcommand{\mimo}{\textsc{mimo}\xspace}

\newcommand{\csi}{\textsc{csi}\xspace}

\newcommand{\stbc}{\textsc{stbc}\xspace}

\newcommand{\snr}{\texttt{SNR}}
\newcommand{\stbcs}{\textsc{stbc}s\xspace}
\newcommand{\ostbc}{\textsc{ostbc}\xspace}
\newcommand{\ostbcs}{\textsc{ostbc}s\xspace}

\newcommand{\BS}{\textsc{bs}\xspace}
\newcommand{\BSs}{\textsc{bs}s\xspace}
\newcommand{\CSI}{\textsc{csi}\xspace}
\newcommand{\DFT}{\textsc{dft}\xspace}

\newcommand{\SNR}{\textsc{snr}\xspace}
\newcommand{\SNRs}{\textsc{snr}s\xspace}

\newcommand{\TDD}{\textsc{tdd}\xspace}

\newcommand{\MIMO}{\textsc{mimo}\xspace} %

\newcommand{\LS}{\textsc{ls}\xspace}

\newcommand{\CDF}{\textsc{cdf}\xspace}
\newcommand{\AWGN}{\textsc{awgn}\xspace}
\newcommand{\LTE}{\textsc{lte}\xspace}

\newcommand{\ID}{\textsc{id}\xspace}
\newcommand{\RF}{\textsc{rf}\xspace}

\newcommand{\tc}{\tau_{\textsc{c}}}
\newcommand{\bpcu}{\text{bpcu}\xspace}

\newcommand{\codeone}{\texttt{1}\xspace}
\newcommand{\codetwo}{\texttt{2}\xspace}
\newcommand{\codefour}{\texttt{4}\xspace}
\newcommand{\codeeight}{\texttt{8}\xspace}
\newcommand{\codetwelve	}{\texttt{12}\xspace}

\newcommand{\boundellipse}[3]
{(#1) ellipse [x radius=#2,y radius=#3]
}

\newcommand{\rate}{R}
\newcommand{\rstar}{R^{*}(n,\epsilon)}
\newcommand{\outagecap}{C_{\epsilon}}
\newcommand{\outageprob}{p_{\text{out}}}
\newcommand{\Reps}{R_{\epsilon}}
\newcommand{\ns}{n_\textsc{s}}

\newcommand{\Xp}{\bX_{\textsc{p}}}
\newcommand{\Xd}{\bX_{\textsc{d}}}

\newcommand{\nt}{n_{\textsc{t}}}

\newcommand{\sn}{s_n}
\newcommand{\rsn}{\bar{s}_n}
\newcommand{\An}{\bA_n}
\newcommand{\Ak}{\bA_k}
\newcommand{\isn}{\tilde{s}_n}
\newcommand{\Bn}{\bB_n}
\newcommand{\Bk}{\bB_k}

\newcommand{\errorone}{\bar{\eta}_1}
\newcommand{\errortwo}{\bar{\eta}_2}
\newcommand{\erroronereal}{\bar{\eta}_1}
\newcommand{\errortworeal}{\bar{\eta}_2}
\newcommand{\erroroneimag}{\tilde{\eta}_1}
\newcommand{\errortwoimag}{\tilde{\eta}_2}

\newcommand{\tp}{\tau_{\textsc{p}}}
\newcommand{\td}{\tau_{\textsc{d}}}

\newcommand{\p}{\rho}
\newcommand{\pp}{\rho_{\textsc{p}}}
\newcommand{\pd}{\rho_{\textsc{d}}}

\newcommand{\byp}{\by_{\textsc{p}}}

\newcommand{\hhat}{\hat{\bh}}

\newcommand{\hhatnorm}{||\hhat||}

\newcommand{\Nb}{N_\text{b}}

\newcommand{\Ch}{\bC_{\bh}}
\newcommand{\Cg}{\bC_{\bg}}

\newcommand{\Ce}{\bC_{\be}}

\newcommand{\dr}{\bPhi}
\newcommand{\dromni}{\bPhi_{\text{\cite{Meng16}}}}
\newcommand{\drrand}{\bPhi_{\textsc{rand}}}
\newcommand{\drdft}{\bPhi_{\textsc{dft}}}
\newcommand{\DRM}{\textsc{drm}\xspace}
\newcommand{\DRMs}{\textsc{drm}s\xspace}
\newcommand{\Xbig}{\boldsymbol{\mathsf{X}}}

\usepackage{siunitx}
\pgfplotscreateplotcyclelist{mylist}{%
{black,mark=o},
{black,mark=triangle},
{black,mark=square},
{black,mark=pentagon},
{black,mark=diamond},
{black,mark=star},}

\newcommand{\systeminfo}{\textsc{si}\xspace}

\newcommand{\xbig}{\boldsymbol{\mathsf{x}}}
\newcommand{\Xpbig}{\boldsymbol{\mathsf{X}}_{\textsc{p}}}
\newcommand{\Es}{E_s}

\newcommand{\hcont}{\bh_{\Sigma}}

\pgfplotsset{StyleCodeOne/.style={%
        solid,mark=o}}
\pgfplotsset{StyleCodeTwo/.style={%
        solid,mark=triangle}}
\pgfplotsset{StyleCodeFour/.style={%
        solid,mark=square}}
\pgfplotsset{StyleCodeEight/.style={%
        solid,mark=diamond, mark options=solid}}
\pgfplotsset{StyleCodeTwelve/.style={%
        solid,mark=pentagon}}

\title{Performance of In-band Transmission of System Information in Massive MIMO Systems}
\author{Marcus Karlsson, Emil Bj{\"o}rnson and Erik G. Larsson
\thanks{The authors are with the Department of Electrical Engineering (ISY), Link{\"o}ping University, 581 83 Link{\"o}ping, Sweden (email: \{marcus.karlsson, emil.bjornson,  erik.g.larsson\}@liu.se).} 
\thanks{This work was supported in part by the Swedish Research Council (VR), and ELLIIT.}
\thanks{A preliminary version of parts of this work was presented at the International Symposium on Wireless Communication Systems (ISWCS), 2015 [17].}}

\IEEEoverridecommandlockouts
\begin{document}
\maketitle
\begin{abstract}
We consider transmission of system information in massive \MIMO. This information needs to be reliably delivered to inactive users in the cell without any channel state information at the base station. Downlink transmission entails the use of downlink pilots and a special type of precoding that aims to reduce the dimension of the downlink channel and the pilot overhead, which would otherwise scale with the number of base station antennas. We consider a scenario in which the base station transmits over a small number of coherence intervals, providing little time/frequency diversity. The system information is transmitted with orthogonal space-time block codes to increase reliability and performance is measured using outage rates. Several different codes are compared, both for spatially correlated and uncorrelated channels and for varying amount of time/frequency diversity. We show that a massive \MIMO base station can outperform a single-antenna base station in all considered scenarios.
\end{abstract}
\section{Introduction}
Massive \MIMO (Multiple-Input Multiple-Output) can bring impressive gains in spectral efficiency, quality of service and fairness compared with contemporary wireless communication systems \cite{Marzetta16,Gao15,Harris15}. Advanced testbeds \cite{Shepard12,Vieira14,Harris15} are already confirming that the theoretical gains and benefits of massive \MIMO can be reaped in practical settings. However, there are still significant problems that need to be solved in order to make it the key technology of the next generation cellular networks. In particular, the base station (\BS) needs some way to convey information about cell operation, such as carrier frequencies, bandwidths, and configurations---commonly called system information (\systeminfo)---to the terminals in the cell. This transmission of \systeminfo is needed for initial access---when an inactive terminal joins the network---and for handover operations. Many papers focus on analyzing the benefits of the technology in the physical layer when the terminals have already received the \systeminfo and are regularly transmitting uplink pilots. Conveying \systeminfo in massive \MIMO has been considered a problem by many in the community and even a show-stopper by some \cite{Bjornson16a}.

When the \BS has channel state information (\CSI), it is able to perform beamforming to achieve a coherent array gain, effectively increasing the signal-to-noise-ratio (\SNR) at the receiving terminals. This means that, when \CSI is available, more terminals can be reached compared to contemporary single-antenna systems, without increasing the transmit power. However, when the \BS does not have \CSI, this array gain is lost. Consequently, there is a gap in the received signal power between the signal carrying \systeminfo, transmitted without \CSI, and the stronger user-dedicated signal, transmitted with \CSI. As a result, the area the \BS can cover without \CSI is smaller than the area covered with \CSI.

A space-time block code (\stbc) can improve the reliability of transmission without \CSI by increasing the effective \SNR at the receiver and by providing spatial diversity. Many contemporary systems use \stbcs, but massive \MIMO offers more freedom in choosing a code because of the larger number of antennas. One specific choice of a \stbc is called \emph{beam sweeping} \cite{Shepard15}, where the \BS sweeps over the cell with the same message using different beams in order to find the terminal. The more antennas the \BS has, the narrower beams it can use, resulting in a high received \SNR whenever the beam ``hits'' the terminal. However, beam sweeping is essentially a spatial repetition code---hence, inefficient.

In this paper, we mainly consider scenarios with stringent latency constraints, high reliability requirements, and a channel that offers little or no time/frequency diversity. A representative scenario could be a narrow-band channel in a cellular system handling \systeminfo or a sensor network using low-energy, narrow-band sensors. We consider using an orthogonal \stbc (\ostbc) which enables full diversity and simple decoding, both desirable in the above-mentioned scenarios. To enable downlink training, a precoding matrix is used to reduce the pilot overhead. Moreover, only in-band solutions are considered, for which the \systeminfo is transmitted in the same frequency band as the payload data.

\subsection{Related Work and Contributions}

Transmission of \systeminfo in massive \MIMO has been considered in \cite{Meng16,Meng16a,Xia16,Qiao16} and is of concern to the industry \cite{Ericsson17}. Reference \cite{Meng16} presents the need for a precoding matrix to reduce pilot overhead, and focuses on optimizing this precoding matrix, constructed from Zadoff-Chu sequences, to achieve approximate omnidirectional transmission on the average. Here, approximate means that the signal powers in all of the $M$ equally spaced discrete angles are identical. The article also measures system performance in terms of the peak-to-average-power ratio of the transmitted signal, outage probability, and ergodic rate when the user has perfect \CSI. In \cite{Meng16a}, the same authors design the \stbc and the precoding matrix jointly, to achieve approximate omnidirectional transmission in each channel use. In \cite{Qiao16,Xia16} omnidirectional transmission, where signal power is constant for \emph{any} angle, not just discrete ones, is considered. The design in \cite{Qiao16} allows for small fluctuations in average power over the angles, while \cite{Xia16} considers omnidirectional transmission, averaged over a few channel uses. Any of these methods regarding \systeminfo can be used together with the method proposed in \cite{Larsson16}, where \systeminfo is transmitted in the same time-frequency resource as the payload data but is confined to the nullspace of the beamforming matrix used for the payload data.

Note that, although all users in the cell receive the same message from the \BS, there is a clear distinction between transmitting \systeminfo and multi-casting in massive \MIMO. When multi-casting \cite{Xiang14}, the \BS exploits \CSI in order to beamform the common information to the terminals. There are also some minor similarities with reducing the dimension of the channel, as done in this paper, and what is known as hybrid beamforming \cite{Molisch16}, where the \BS uses a low dimensional digital precoder and maps the output of this to the antenna array with a high dimensional analogue precoder, consisting of phase shifters. Some prominent differences between hybrid beamforming and dimension reduction are: hybrid beamforming is limited by the number of \RF chains and their resolution, but in this paper, each antenna has its own \RF chain; many of the algorithms used in hybrid beamforming aim to maximize the spectral efficiency, ignoring the users with poor channel conditions; and hybrid beamforming needs \CSI which is not available to the \BS in the considered scenario. Additionally, there is no guarantee that the dimension reduction with a given \stbc can be realized using hybrid beamforming.

The specific contributions of the paper are the following:
\begin{itemize}

\item We derive a lower bound on the \SNR obtained at the terminal for downlink communication in a massive \MIMO system using downlink pilots and an arbitrary \ostbc without any prior \CSI available to the terminal or the \BS. This bound is found to be close to a bound that follows as a special case of the results in \cite{Larsson16}, where no structure of the transmitted signal is assumed.

\item We analyze the need for spatial diversity for transmission of \systeminfo in a massive \MIMO system by comparing the performance of several \ostbcs in correlated and uncorrelated channels. For the considered scenario, using codes providing a higher diversity order than around $10$ is not beneficial. For larger codes the increase in spatial diversity is not enough to counteract the pre-log penalty associated with the pilot overhead.

\item We study how the availability of time-frequency resources for \systeminfo affects the choice of \ostbcs. Here we consider two cases: First, the amount of information the \BS wants to convey to the terminal is fixed and the \BS minimizes the amount of time-frequency resources used. Second, the amount of time-frequency resources available for \systeminfo is fixed and the \BS aims to convey as much information as possible to the terminal.

\item We derive a corresponding lower bound on the \SNR at the terminal, for the case of a multi-cell system with different pilot reuse, and compare performance to that of the single-cell system.
\end{itemize}

In earlier conference papers we have presented some initial results. In \cite{Karlsson14a}, we highlighted the need for downlink pilots for transmission without \CSI at the \BS and introduced the idea of spatially repeating a small code over the antennas. Reference \cite{Karlsson15} treated a scenario similar to the one in the current paper, but the analysis here is includes correlated channels, larger and rectangular \ostbcs, least-squares (\LS) estimation, pilot-energy optimization, and multiple cells.

\textbf{Notation:} Boldfaced lowercase letters, $\bx$, denote column vectors, boldface uppercase letters, $\bX$, denote matrices and lower case letters, $x$, denote scalars. $\bI_{M}$ is the identity matrix of dimension $M \times M$ and $\bzero_{a\times b}$ is the zero matrix of dimensions $a\times b$. $\bX^{*}$, $\bX\trans$ and $\bX\herm$ denote conjugate, transpose and Hermitian transpose, respectively. The 2-norm of a vector $\bx$ is denoted by $\norm{\bx}$. $\bar{x}\triangleq\Re(x)$ and $\tilde{x}\triangleq\Im(x)$ denote the real and imaginary parts, respectively, and the imaginary unit is denoted by $\iu$. $\CN(\bx,\bX)$ represents the circularly symmetric, complex Gaussian distribution with mean $\bx$ and covariance matrix $\bX$ and $\chi^{2}(m)$ is a Chi-squared distribution with $m$ degrees of freedom. The notation $f(x) = \ordo(g(x))$ means that there exist positive constants $c$ and $x_0$ such that 
\[ |f(x)|\leq c |g(x)|,\ \forall x \geq x_0. \]

\section{Background}\label{sec:system_model}

\subsection{Orthogonal Space-Time Block Codes}
This subsection introduces \ostbcs and their associated terminology, starting with the more general linear \stbcs. The information in this section can be found in, for example \cite{Larsson03}, but some key equations are stated here in order to make the paper self-contained as well as to establish notation and terminology.
 
A linear \stbc is a code for which each code matrix (codeword) $\bX$ carries $\ns$ information bearing symbols over $\tau$ channel uses, using $\nt$ antennas. That is, each code matrix $\bX$ is a $\tau\times\nt$ (complex-valued) matrix of the form
\begin{equation}\label{eq:lstbc}
\bX = \sum\limits_{n=1}^{\ns} \rsn\An + \iu\isn\Bn,
\end{equation}
where $\rsn$ ($\isn$) is the real (imaginary) part of the symbol to be transmitted, $\sn = \rsn + \iu\isn$. $\An$ and $\Bn$ are fixed $\tau\times\nt$, generally complex-valued, matrices which define the specific code. Since $\ns$ symbols are conveyed over $\tau$ channel uses, the \emph{code rate} is $\ns/\tau$. We also refer to $\tau$ as the decoding delay, or simply \emph{delay}, since the receiver has to wait $\tau$ channel uses before decoding the codeword $\bX$. We will further refer to $\nt$ as the \emph{size} of the code. Specifically, a ``larger code'' means a code with larger $\nt$.

An \ostbc is a linear \stbc for which all code matrices satisfy \[ \bX\herm\bX = \sum\limits_{n=1}^{\ns} |\sn|^{2}\bI_{\nt}.\]
This implies that $\tau\geq\nt$. This orthogonality also means that the symbols decouple in coherent detection \cite[Section 7.4]{Larsson03}, \cite{Tarokh99a}.

All \ostbcs satisfy the following identities \cite[Theorem 7.1]{Larsson03}:
\begin{equation*}
\begin{aligned}
\An\herm\An &= \bI_{\nt},\ \Bn\herm\Bn = \bI_{\nt},&\\
\An\herm\Ak &= - \Ak\herm\An,\ \Bn\herm\Bk = -\Bk\herm\Bn, &\forall n,k,n\neq k,\\
\An\herm\Bk &= \Bk\herm\An, &\forall n, k.
\end{aligned}
\end{equation*} 
From these identities one can deduce that for any complex-valued vector $\bv$
\begin{equation}\label{eq:ostbcidentity:AnAp}
\Re(\bv\herm\An\herm\Ak\bv) = 
\begin{cases}
0, &n\neq k\\
\norm{\bv}^{2}, &n = k
\end{cases}
\end{equation}
and
\begin{equation}\label{eq:ostbcidentity:BnAp}
\Re(-\iu\bv\herm\An\herm\Bk\bv) = \Im(\bv\herm\An\herm\Bk\bv) =  0, \forall n,k,
\end{equation}
which will prove useful later.

As a special case of \Eqref{eq:lstbc}, consider letting $\sn = s$ for $n=1,\dots,\ns$, then
\[ \bX = \bC s,  \]
for some complex matrix $\bC$. This is one way of describing beam sweeping, where the rows of $\bC$ are designed to provide spatial coverage. We see here that beam sweeping is a special case of a linear \stbc with code rate $1/\tau$.

In this paper, we consider four different \ostbcs. As a reference, we also consider a \BS with a single antenna. The considered \ostbcs are listed and summarized in \Tableref{tab:codes}. When referring to the codes, we will use the code identity (\ID), defined in \Tableref{tab:codes}. Code \codetwo is the Alamouti code \cite{Alamouti98} and code \codefour can be found in \cite{Larsson03}. Codes \codeeight and \codetwelve were created following the algorithm outlined in \cite{Tarokh99a}.
\begin{table}
	\centering
	\caption{The parameters of the \ostbcs considered in the paper.}\label{tab:codes}
	\begin{tabular}{|c|c|c|c|c|}
		\hline
		Code \ID		& $\nt$	& $\td$		& $\ns$	& Code Rate $\left(\dfrac{\ns}{\td}\right)$ \\\hline
		\codeone		& $1$		& $1$		& $1$		& $1$ \\\hline
		\codetwo		& $2$		& $2$		& $2$		& $1$ \\\hline
		\codefour		& $4$		& $4$		& $3$		& $3/4$\\\hline
		\codeeight		& $8$		& $16$		& $8$		& $1/2$\\\hline
		\codetwelve		& $12$		& $128$		& $64$		& $1/2$\\\hline
	\end{tabular}
\end{table}

The two larger codes in \Tableref{tab:codes} are suboptimal, both in terms of rate \cite{Liang03} and delay \cite{Das12}. This guarantees that an optimal code (in terms of rate, delay, or both) will perform at least as well. The main point, however, is that a massive \MIMO \BS can outperform a single-antenna \BS and to show this, the codes in \Tableref{tab:codes} are more than enough.

\subsection{The Finite Coherence Interval}

The coherence interval is a time-frequency block whose time-duration is equal to the coherence time and whose bandwidth is equal to the coherence bandwidth. The size of the coherence interval in samples, denoted $\tc$, can vary vastly between applications, from a few hundred symbols, to practically infinite \cite[Chapter 2]{Marzetta16}. For an inactive user, the \BS does not know the length of the coherence interval, and hence has to use a conservative estimate in order to reduce the risk of overestimating the stability of the channel. In practice, the system is limited by the channel offering the smallest coherence interval.

The finite coherence interval is the reason why massive \MIMO requires time-division duplex (\TDD) operation in order to be scalable in the number of \BS antennas, unless additional assumptions on propagation are made \cite{Bjornson16a}. \TDD enables channel reciprocity within a coherence interval, which allows the \BS to learn the uplink and downlink channels from uplink pilots. If downlink pilots were used, a \BS with $M$ antennas would have to spend at least $M$ channel uses on downlink training in every coherence interval, plus additional feedback.

\section{System Model}\label{sec:systemmodel}

The paper will focus on the single-cell case, where no interference from other cells is present, as most of the interesting phenomena arise there. However, we will provide a brief discussion of what changes in a multi-cell scenario in \Secref{sec:snr:multicell} and compare some of the results for the single-cell scenario to that for the multi-cell scenario.

Consider a single-cell system in which the \BS is equipped with $M$ antennas and wishes to convey \systeminfo to an arbitrary single-antenna user within the cell. Neither the \BS nor the terminal has any a priori \csi. The received signal at the terminal is
\[ y \triangleq \sqrt{\p}\xbig\trans\bg + w, \]
where $\xbig\in\mC^{M\times 1}$, $\bg\in\mC^{M\times 1}$, and $w$ are the transmitted signal, the channel, and noise, respectively. The transmitted signal $\xbig$ satisfies $\E [\xbig\herm\xbig] = 1$, $\p$ is the normalized transmit power and $w\sim \CN(0,1)$ is independent, normalized noise. The channel $\bg$ is assumed to be distributed as $\CN(\bzero,\Cg)$, where $\Cg\triangleq\Exp{\bg\bg\herm}\in\mC^{M\times M}$ is the channel covariance matrix. Over $\tau$ channel uses the \BS transmits the $\tau\times M$ matrix 
\[ \Xbig \triangleq 
\left[\begin{array}{c}
\xbig_1\trans\\
\xbig_2\trans\\
\vdots\\
\xbig_{\tau}\trans
\end{array}\right], \]
whereby the user receives the $\tau\times 1$ vector
\begin{equation}\label{eq:receivedsignal}
\by \triangleq \sqrt{\rho}\Xbig\bg + \bw,
\end{equation}
where 
\[ \bw \triangleq [w_1,\dots,w_\tau]\trans\]
has independent $\CN(0,1)$ elements.

When the user detects the transmitted symbols, it is beneficial to have \CSI. To give the terminal \CSI, the \BS first transmits the pilot matrix  $\Xpbig\in\mC^{\tp\times M}$, known a priori to both parties. Orthogonal pilots ($\Xpbig\herm\Xpbig\propto\bI_{M}$) are usually preferred as they are optimal in a mean square error sense \cite[Section 9.4]{Larsson03} in independent, identically distributed (i.i.d.) Rayleigh fading. Additionally,  orthogonal pilots ensure that the channel coefficients decouple during estimation in i.i.d.~Rayleigh fading. However, transmitting orthogonal pilots would require $\tp \geq M$, which means spending many channel uses on pilots. If $\tp$ is of the same order as the coherence interval $\tc$, few channel uses will be left for data, and if $\tp > \tc$, the orthogonal pilot block would be too big to fit in one coherence interval.

Transmitting \systeminfo is seemingly the only time, apart from a computational complexity perspective, when a massive \MIMO system does not benefit from having more antennas. If the \BS only had a few antennas, there would be no problem sending orthogonal downlink pilots. To resolve this problem, there are a few alternatives: i) Restrict the number of antennas at the \BS for the sole purpose of being able to transmit orthogonal downlink pilots when conveying \systeminfo. This is not an appealing solution since it eliminates the benefits of massive \MIMO. ii) Turn off antennas and transmit \systeminfo on only a subset of the array. This is problematic because either the transmission without \CSI will have to be done with a fraction of the total output power used in coherent transmission, or the hardware has to be able to work with large variations in output power, which would make the hardware more expensive. iii) Make use of the excess of degrees of freedom and spatial diversity, provided by the abundance of antennas at the \BS. iv) Use a single, more powerful antenna operating at another frequency, dedicated to provide \systeminfo. As this paper only considers in-band solutions, option iv is out of scope.

We consider the third alternative, and aim to find a middle ground between full repetition over the antennas (beam sweeping), associated with a lower rate, and no repetition, associated with a large pilot overhead.

\subsection{The Dimension Reducing Matrix}

As mentioned earlier in \Secref{sec:systemmodel}, having a \BS with a moderate or even small number of antennas might be beneficial, considering the same total output power. To emulate a \BS with few antennas, consider constructing the transmitted signal $\Xbig$ with two separate parts:
\begin{equation}\label{eq:transmitted:signal}
\Xbig= \bX\dr ,
\end{equation}
where $\dr\in\mC^{\nt\times M}$, $\nt <  M$ is a (deterministic) precoding matrix called the \emph{dimension-reducing matrix} (\DRM), with the purpose of spreading the \ostbc $\bX\in\mC^{\tau\times\nt}$ over the antennas. With \Eqref{eq:transmitted:signal}, the received signal \Eqref{eq:receivedsignal} can be written as
\[ \by = \sqrt{\rho}\bX\dr\bg + \bw = \sqrt{\rho}\bX\bh + \bw,\]
where we have defined the \emph{effective channel} $\bh\triangleq\dr\bg\in\mC^{\nt}$. The \DRM effectively shrinks the channel dimension from $ M$ to $\nt$. The matrix $\bX$ can be thought of as the output of $\nt$ antenna ports, and $\dr$ represents the mapping from the antenna ports to the physical antennas. 

After choosing a \DRM, the \BS can transmit \systeminfo to the users in the cell over the effective channel $\bh$. The transmission is divided into the pilot phase, in which the \BS transmits a predetermined set of pilots in the downlink, and the data phase, in which information-bearing symbols are transmitted. Note that the \DRM has to remain constant for the entire coherence interval, i.e., over both the pilot and the data phase. 

\subsection{Pilot Phase}\label{sec:pilot:phase}

As long as the \DRM is fixed within a coherence interval, the effective channel $\bh$ is static, which means it can be estimated. To estimate the channel, a semi-unitary pilot matrix $\Xp\in\mC^{\tp\times\nt}$, $\tp \geq \nt$, satisfying 
\[ \Xp\herm\Xp = \dfrac{\tp}{\nt}\bI_{\nt}\] 
is transmitted with normalized transmit power $\pp$ by the \BS. The received signal at the terminal is
\[ \byp \triangleq \sqrt{\pp}\Xp\bh + \bw. \]
Since the terminal lacks \CSI, the \LS estimate of the channel is used:
\begin{equation}\label{eq:ls:estimate}
\hhat \triangleq (\sqrt{\pp}\Xp\herm\Xp)\inv\Xp\herm\byp
=
(\Xp\herm\Xp)\inv\Xp\herm\Xp\bh + (\sqrt{\pp}\Xp\herm\Xp)\inv\Xp\herm\bw
=\bh + \be,
\end{equation}
where
\[ \be \triangleq (\sqrt{\pp}\Xp\herm\Xp)\inv\Xp\herm\bw = \hhat - \bh \]
is the channel estimation error. The channel and the estimation error have covariance matrices
\begin{equation}\label{eq:effective:covariance}
\Ch \triangleq \Exp{\bh\bh\herm} = \Exp{\dr\bg\bg\herm\dr\herm}
= \dr\Cg\dr\herm
\end{equation}
and 
\begin{equation}\label{eq:effective:covariance:error}
\Ce \triangleq \Exp{\be\be\herm} = \dfrac{\nt}{\pp\tp}\bI_{\nt},
\end{equation}
respectively.

The channel estimate $\hhat$ and the channel estimation error $\be$ are jointly Gaussian and correlated. From \cite[Theorem 10.2]{Kay93} we can write
\[ \be|\hhat \sim \CN(\bU\hhat, \bR), \]
where 
\[ \bU \triangleq \Ce\left(\Ce + \Ch\right)\inv, \text{ and } \bR \triangleq \left(\Ce\inv + \Ch\inv \right)\inv. \]
In particular, this means that
\begin{equation}\label{eq:conditional:moments}
\Exp{\be\given \hhat} = \bU\hhat,\
\Exp{\be\be\herm\given \hhat} = \bR + \bU\hhat\hhat\herm\bU\herm,\
\Exp{\be\be\trans\given \hhat} = \bU\hhat\hhat\trans\bU\trans.
\end{equation}

\subsection{Data Phase}\label{sec:transmission:data}

In the data phase, the \BS transmits \ostbc matrices $\Xd\in\mC^{\td\times\nt}$, conveying $\ns$ mutually independent information-bearing symbols over $\td$ channel uses. With $\pd$ denoting the normalized transmit power, the received signal at the terminal is
\[ \by = \sqrt{\pd} \Xd\bh  + \bw.\] 
The codeword $\Xd$ satisfies 
\[ \Exp{\trace (\Xd\herm\Xd)} = \td, \]
and the symbol energy is 
\[ \Es \triangleq \Exp{|s_n|^{2}} = \dfrac{\td}{\ns\nt}.\]

In order to detect the complex symbol $\sn$, the user treats the estimated channel $\hhat$ as the true channel and detects the real and imaginary part of $\sn$ separately. To detect the real part of the transmitted symbol, $\rsn$, the terminal multiplies the received vector with $\hhat\herm\An\herm$ from the left and takes the real part\cite{Larsson03}:
\begin{equation}\label{eq:detect:real}
\hat{\bar{s}}_n 
= \real{\hhat\herm\An\herm\by}
= \real{\sqrt{\pd}\hhat\herm\An\herm\Xd\hhat}
- \real{ \sqrt{\pd}\hhat\herm\An\herm\Xd\be}
+ \real{\hhat\herm\An\herm\bw}.
\end{equation}
From \Eqref{eq:ostbcidentity:AnAp} and \Eqref{eq:ostbcidentity:BnAp}, 
\[ \real{\sqrt{\pd}\hhat\herm\An\herm\Xd\hhat}
=\sqrt{\pd} \hhatnorm^{2}\rsn.\]
The last two terms in \Eqref{eq:detect:real} are denoted by
\[ \errorone \triangleq -\real{ \sqrt{\pd}\hhat\herm\An\herm\Xd\be}\]
and 
\[ \errortwo \triangleq \real{\hhat\herm\An\herm\bw}. \]
We can now write the received, processed, real symbol as
\[\hat{\bar{s}}_n =  \sqrt{\pd} \hhatnorm^{2}\rsn + \errorone + \errortwo.\]

To decode the imaginary part of $\sn$, we use $-\iu\hhat\herm\Bn\herm$ instead of $\hhat\herm\An\herm$ and the following calculations are otherwise identical to what we have above. This calculation gives the error terms 
\[ \erroroneimag 
\triangleq 
-\real{ -\iu\sqrt{\pd}\hhat\herm\Bn\herm\Xd\be}
=
-\imag{\sqrt{\pd}\hhat\herm\Bn\herm\Xd\be}\]
and 
\[ \errortwoimag 
\triangleq 
\real{-\iu\hhat\herm\Bn\herm\bw}
=
\imag{\hhat\herm\Bn\herm\bw}, \]
completely analogous to $\erroronereal$ and $\errortworeal$ for the detection of the real part. Finally, we can write the received, processed complex symbol as
\begin{equation}\label{eq:detect:complex}
\hat{s}_n
\triangleq  
\sqrt{\pd}\hhatnorm^{2}(\rsn+\iu\isn) + \errorone + \iu\erroroneimag + \errortwo + \iu\errortwoimag
= 
\sqrt{\pd}\hhatnorm^{2}s_n + \eta_1+ \eta_2,
\end{equation}
where $\eta_1 \triangleq \errorone + \iu\erroroneimag$ and $\eta_2 \triangleq \errortwo + \iu\errortwoimag$.

Conditioned on the channel estimate, $\hhat$, \Eqref{eq:detect:complex} is a deterministic channel plus noise. The first error term $\eta_1$, stemming from the imperfect channel estimate, is correlated with the symbol of interest $\sn$. We can thus write
\[ \eta_1 = c_n\sn + u_n, \]
where $c_n \triangleq \Exp{\sn\conj\eta_1\given\hhat}/\Es$ and $u_n$ is uncorrelated to $\sn$. With this, \Eqref{eq:detect:complex} becomes
\begin{equation}\label{eq:complex:symbol:uncorrelated}
\hat{s}_n = \left(\sqrt{\pd}\hhatnorm^{2} + c_n\right)s_n + u_n + \eta_2.
\end{equation}
The signal in \Eqref{eq:complex:symbol:uncorrelated} is now uncorrelated to the noise, conditioned on $\hhat$, and the received \SNR is given by \cite{Medard00}
\begin{equation}\label{eq:bound}
\dfrac{\left|\Exp{\sn\conj\hat{s}_n\given \hhat}\right|^{2}}{\Es\Exp{\left|\hat{s}_n\right|^{2}\given\hhat} - \left|\Exp{\sn\conj\hat{s}_n\given\hhat}\right|^{2}}.
\end{equation}

With 
\[ U_n \triangleq \Exp{|u_n|^{2}\given\hhat} = \Exp{|\eta_1|^{2}\given\hhat} - \Es\Exp{|c_n|^{2}\given\hhat} \]
and
\[ \Exp{|\eta_2|^{2}\given \hhat} = \norm{\hhat}^{2}, \]
\Eqref{eq:bound} can be expressed as 
\begin{equation}\label{eq:snr:symbol}
\snr_n \triangleq \frac{\Es\left|\sqrt{\pd}\hhatnorm^{2} + c_n \right|^{2}}{U_n + \hhatnorm^{2}}. 
\end{equation}
Note that the \SNR in \Eqref{eq:snr:symbol} can vary between symbols for the same channel realization. This variation in \SNR is small: in the order of $0.1$ percent for all analyzed cases. We define the achievable \SNR when using an \ostbc as
\begin{equation}\label{eq:snr:ostbc}
\snr^{\ostbc} \triangleq \min_{n\in\{1,\dots,\ns\}} \snr_n.
\end{equation}
In the special case when the physical channel has i.i.d.~elements,
\[ \Cg = \bI \beta, \]
where $\beta$ represents the large-scale fading, we have
\[ c_n = -\sqrt{\pd}\hhatnorm^{2}\dfrac{\nt}{\beta\tp\pp+\nt}. \] If, in addition, the code is a square \ostbc ($\nt = \td$)
\[ U_n =  \dfrac{\pd{\td\beta}}{\beta\tp\pp + \nt}\hhatnorm^{2}\] and the symbol \SNR in \Eqref{eq:snr:symbol} can be simplified to
\begin{equation}\label{eq:snr:square}
\snr^{\text{square}} \triangleq
\dfrac{\Es\pd{\hhatnorm^{2}}}{\dfrac{\pd\td\beta}{\nt+\beta\pp\tp}+1}\left(\dfrac{\beta\tp\pp}{\beta\tp\pp+\nt}\right)^{2}.
\end{equation}

We will later numerically compare the outage rate achieved when using \Eqref{eq:snr:ostbc} to the rate achieved when using the \SNR derived in \cite[Eq. (49)]{Larsson16}, where no structure of the transmitted signal was assumed. The \SNR from \cite{Larsson16} is given by\footnote{In \cite{Larsson16}, the data power and the pilot power are assumed to be equal, which is not the case here. In addition, we do not consider simultaneous payload transmission, so $\rho_{b}'$ is zero. The \SNR expression has been modified accordingly.} 
\begin{equation}\label{eq:snr:general}
\snr^{\text{general}} \triangleq \dfrac{\dfrac{\pd}{\nt}\norm{\hhat_{\textsc{mmse}}}^{2}}{ \dfrac{\nt\pd\beta}{\nt+\tp\pp\beta} + 1},
\end{equation}
where $\hhat_{\textsc{mmse}}$ is the channel estimate if a minimum-mean-square-error estimator is used by the terminal. The \SNR in \Eqref{eq:snr:general} can be seen as an upper bound on the \SNR in \Eqref{eq:snr:ostbc}, as the former does not assume any structure of the transmitted signal.

When a square \ostbc with full rate ($\ns=\nt=\td$) is used, \Eqref{eq:snr:square} and \Eqref{eq:snr:general} are distributed identically as
\[ \snr^{\text{square}} \sim \snr^{\text{general}} \sim \dfrac{\pp\tp\pd\td\beta^{2}}{2\ns\nt\left(\pd\td\beta + \pp\tp\beta + \nt\right)}\chi^{2}(2\nt). \]
This can be shown by observing that
\[ \hhatnorm^{2}\sim \dfrac{\nt + \pp\tp\beta}{2\pp\tp}\chi^{2}(2\nt)  \]
and
\[ \norm{\hhat_{\textsc{mmse}}}^{2}\sim \dfrac{\pp\tp\beta^{2}}{2(\pp\tp\beta + \nt)}\chi^{2}(2\nt).  \]

\subsection{The Multi-Cell Scenario}\label{sec:snr:multicell}

Deriving the lower bound on the \SNR for the multi-cell case follows a similar route as in the single-cell case, only with more terms. We let $K$ denote the number of interfering cells and $\mK$ denote the set of \emph{contaminating} cells that use the same pilots as the home cell. The pilot sequences used by cells not in $\mK$ are orthogonal to the pilot sequence used in the home cell. For a pilot reuse of $p$, at least $p\nt$ channel uses will be occupied by pilots.

\subsubsection{Pilot Phase} Following the same steps as in \Secref{sec:pilot:phase}, the multi-cell equivalent to the channel estimate $\hhat$ can be written as
\[ \hhat_{\textsc{mc}} = \bh + \be + \hcont, \]
where 
\[ \hcont \triangleq \sum\limits_{k\in\mK}\bh_k \]
and $\bh_k$ is the channel from the \BS in cell $k$ to the terminal in the home cell. Just as in the single-cell case, the estimation error, now $\be + \hcont$, is correlated to the channel estimate. 

To calculate \Eqref{eq:bound}, the conditional moments of $\bh_k|\hhat_{\textsc{mc}}$ for $k=1,\dots,K$, $\hcont|\hhat_{\textsc{mc}}$, and $\be|\hhat_{\textsc{mc}}$ are needed, as well as the conditional moments of $\be|\hhat_{\textsc{mc}},\hcont$. These can be found by using Bayes' theorem.

\subsubsection{Data Phase} When detecting the information-bearing symbol, two additional noise terms show up, compared to the single-cell case:
\[ \eta_3 \triangleq -\sqrt{\pd}\real{\hhat_{\textsc{mc}}\herm\An\herm\Xd\hcont} -\iu\sqrt{\pd}\imag{\hhat_{\textsc{mc}}\herm\Bn\herm\Xd\hcont} \]
and 
\[ \eta_4 \triangleq \sum\limits_{k=1}^{K}\real{\hhat_{\textsc{mc}}\herm\An\herm\bX_k\bh_k} + \iu\imag{\hhat_{\textsc{mc}}\herm\Bn\herm\bX_k\bh_k}, \]
where $\bX_k$ is the signal transmitted from cell $k$ in the data phase. All cells are assumed to transmit data in the same time-frequency resource. This gives an expression for the received, processed signal in a multi-cell scenario
\[ \hat{s}_n = \sqrt{\pd}\norm{\hhat_{\textsc{mc}}}^{2}\sn + \eta_1 + \eta_2 + \eta_3 + \eta_4. \]
Note that $\eta_1$ and $\eta_3$  are correlated, both to each other, and to the symbol $\sn$ and that $\eta_2$ and $\eta_4$ are uncorrelated to all other terms. To calculate the \SNR in \Eqref{eq:bound}, one can split $\eta_1$ and $\eta_3$ into parts that correlate perfectly with $\sn$, and a part that is uncorrelated to $\sn$, as done in \Secref{sec:transmission:data}.

\subsection{OSTBCs in Massive MIMO}

Because a massive \mimo \BS has an abundance of transmit antennas, it generally has more options in the signal design compared to contemporary \BSs. For example, the \BS has, to a greater extent, the ability to  dynamically change what \stbc to use. If the \BS is equipped with $M$ antennas, the size of the code (number of antenna ports) $\nt$ can be changed to suite the scenario in question. If high reliability is needed, and there is little time/frequency diversity in the channel, the \BS can choose a large $\nt$ to compensate the lack of time/frequency diversity by adding spatial diversity. If the channel offers enough time/frequency diversity, a code with low diversity and high rate may be used. The caveat here is, as we will see, that even if the \BS may choose $\nt$ to be any integer between 1 and $ M$ in theory, a very large value of $\nt$ is not possible or useful in practice.

There are limits to how high rate an \ostbc spanning $\nt$ antenna ports can have. For example, no \ostbc can have a rate higher than $1$, and for $\nt>1$, this rate is only achievable with $\nt = 2$ (the Alamouti code). The maximum rate of an \ostbc with $\nt = 2m$ or $\nt = 2m-1$, with $m$ being an integer is $\frac{m+1}{2m}$. In particular, as $\nt$ grows, the maximum rate approaches $1/2$\cite{Liang03}. 

The second dimension of an \ostbc, the delay $\td$, becomes more important the larger $\nt$ is as $\td\le \tc-\tp$ is required for the code to fit into one coherence interval. In general, for a fixed code rate $\ns/\td$, delay increases quite fast with $\nt$. The minimum delay grows especially fast when \ostbcs with optimal rate are considered. For example, the minimum delay of a maximum rate code with $\nt = 8$ antennas is $\td=56$ channel uses, and for a code with $\nt = 16$, the minimum delay is $11440$ channel uses \cite{Adams07}.

Hence, we have a practical limit to the code size $\nt$. The limiting factor for massive \MIMO, when it comes to choosing an \ostbc is the decoding delay together with the finite coherence interval, not the number of antenna ports. This means that increasing $\nt$ stops being useful at some point, since the decoding delay is too long.
 
\section{Impact of the Dimension Reducing Matrix}

Let us now consider the transmission over the effective channel $\bh$. The statistics of $\bh$ depend on the choice of the \DRM $\dr$ and the statistics of the physical channel $\bg$ as indicated by \Eqref{eq:effective:covariance}. Apart from studying i.i.d.~Rayleigh fading we also consider a correlated channel model which is described in \Secref{sec:covariance}. The choice of \DRM and how the channel statistics affect this choice is discussed in \Secref{sec:choosing:drm}.

\subsection{Channel Covariance Matrix}\label{sec:covariance}

To understand how correlation between antennas affects performance, we model the correlation of the antenna array with an exponential correlation matrix \cite{Loyka01}. This model has the beauty of being parameterized by a single complex parameter, $r$, denoting the (complex) correlation between the channels of two neighboring antennas. The $(i,j)$:th element of the covariance matrix $\Cg$ is given by
\[\Cg(i,j) =\beta|r| ^{|j-i|}e^{\iu\arg(r)(j-i)}\]
with $|r| \leq 1$. This means that channels for antennas further apart have a smaller correlation, which is physically reasonable. 

Two interesting special cases of this correlation matrix happen when $r = 0$ or $|r| = 1$. For $r=0$, $\Cg$ is a scaled identity matrix and hence corresponds to i.i.d. fading. If $|r| = 1$, then all columns of $\Cg$ are linearly dependent, so the correlation matrix has rank 1. Note that for large arrays, even when $|r|$ is close to 1, the correlation between antennas at moderate distance becomes negligible, as the correlation decays exponentially with the antenna distance.

The complex parameter $r=|r|e^{\iu\arg(r)}$ depends on the magnitude $|r|$ and the argument $\arg (r)$. In the numerical results, we fix $|r|\in [0,1]$ and let $\arg (r)$ vary depending on the user position. We set $\arg (r)$ to be the angle of incidence (as if a line-of-sight channel) from the user to the \BS array. This means that we only need to specify $|r|$.

\subsection{Choosing the Dimension-Reducing Matrix}\label{sec:choosing:drm}

Any choice of $\bPhi$ confines the effective channel $\bh$ to the subspace spanned by the columns of $\bPhi$; the \BS implicitly beamforms into this subspace. For physical channels $\bg$ in the approximate nullspace of $\bPhi$, the effective channel gains will be small. There is an intricate connection between the choice of \DRM and the resulting \SNR, since the \DRM $\dr$ shows up at several places in \Eqref{eq:snr:symbol}. The question is how to choose a suitable \DRM depending on, among other things, the chosen code and number of \BS antennas. Note that there is no obvious ``optimal'' \DRM here. One way of finding an upper bound on performance would be to assume perfect \CSI at the \BS; however, in this case, the \BS would be able to beamform in a conventional manner (by for example multi-casting), making the comparison void.

When transmitting \systeminfo, the \BS does not know who is listening; hence the choice of \DRM should not depend on the physical channel $\bg$. However, if the \BS has statistical knowledge of the channel, this could be used when constructing the \DRM. Recall that we do not assume any channel knowledge, statistical or instantaneous, at the \BS.

To illustrate the importance of the \DRM, we compare three different strategies for choosing the \DRM:
\begin{itemize}
\item The first \DRM considered is the one derived in \cite[Eq. (30)]{Meng16}. This matrix, denoted $\dromni$, has several desirable properties: it ensures approximate omnidirectional transmission, equal output power on all antennas on the average and signals with low peak-to-average-power ratio.
\item Second, we choose a random \DRM:
\[ \drrand = \left[\begin{array}{c}
\bI_{\nt}\ \bzero_{\nt\times( M-\nt)}
\end{array}\right]\bQ, \]
where $\bQ\in\mC^{ M\times M}$ is an isotropically distributed unitary matrix \cite{Marzetta99}, in order to make the matrix ``as random as possible''. 
\item Third, we choose the \DRM as $\nt$ evenly spaced columns in the $ M$ dimensional discrete Fourier transform (\DFT) matrix. That is, the columns with indices
\[ \dfrac{ M}{2\nt}(2n-1),\ n=1,\dots,\nt. \]
We denote this matrix by $\drdft$.
\end{itemize}
The second and third choices are heuristic. The \DRM $\drrand$ demonstrates the performance of a matrix without any particular structure. This is a reasonable choice if the \BS has no idea what effect the \DRM has on the transmission. The motivation for $\drdft$ is that the columns of the \DFT matrix corresponds to different angular directions. By spreading out the angles, at least one of them should work reasonably well for any given terminal. We expect $\dromni$ to outperform the other two, as this is optimized. The main reason we present the other two is to show that a seemingly reasonable choice ($\drdft$) can perform poorly, while a random matrix ($\drrand$) can perform well. 

\emph{Remark:} There are minor similarities between the \DRM used here and the prebeamforming matrix used in \cite{Adhikary13}: both matrices can be built up from selected columns of the \DFT matrix and simplify the channel estimation. However, the prebeamforming matrix has another purpose: to divide known users in the cell into groups based on the eigenspace of the users' covariance matrices. This is a completely different scenario than considered herein, where statistical \CSI is available to the \BS, the channel model is different, and payload data is transmitted.

Note that all three choices of \DRMs are semi-unitary: $\dr\dr\herm=\bI_{\nt}$. For i.i.d.~Rayleigh fading, $\Cg = \beta\bI_{ M}$, this implies that the effective channel $\bh$ will have the same statistics for any choice of \DRM:
\[ \Ch = \beta\dr\bI_{ M}\dr\herm = \beta\bI_{\nt}. \]
Thus, all three choices are equivalent and the choice only makes a difference when $\Cg$ is not a scaled identity matrix.

The cell edge \SNR is defined as the \SNR experienced by a terminal on the cell edge, if all power were transmitted from a single antenna in the array. Throughout the paper, we have a cell edge \SNR of $-5$ \si{\dB}.

\begin{figure}[t]
	\subfloat[]{ %
	\label{fig:snr:drm:two}
	\begin{tikzpicture}
	\begin{axis}[
	xmin=-54,
	xmax=4,
	ymin=0.0001,
	ylabel near ticks,
	ymode = log,
	xlabel near ticks,
	table/y index={1},
	table/x index={0},
	table/col sep=comma,
	xlabel={\SNR [dB]},
	ylabel={\CDF},
	legend pos=south east,
	legend style ={draw=none, fill=none},
	]
	\addplot[black, thick, solid] table {sampled-snr-drm-opt-24-22.dat};
	\addlegendentry{$\dromni$};
	\addplot[black, thick, dashed] table {sampled-snr-drm-best-24-22.dat};
	\addlegendentry{$\drrand$ (best)};
	\addplot[black, very thick, dotted] table {sampled-snr-drm-worst-24-22.dat};
	\addlegendentry{$\drrand$ (worst)};
	\addplot[black, dashdotted, thick] table {sampled-snr-drm-dft-24-22.dat};
	\addlegendentry{$\drdft$};
	\end{axis}
	\end{tikzpicture}}
	\subfloat[]{
	\label{fig:snr:drm:eight}
	\begin{tikzpicture}
		\begin{axis}[
		ymin=0.0001,
		xmin=-54,
		xmax=4,
		ymode = log,
		xlabel near ticks,
		table/y index={1},
		table/x index={0},
		table/col sep=comma,
		xlabel={\SNR [dB]},
		legend pos=north west,
		legend style ={draw=none, fill=none},
		ymajorticks=false,
		]
		\addplot[black, thick, solid] table {sampled-snr-drm-opt-120-816.dat};
		\addlegendentry{$\dromni$};
		\addplot[black, thick, dashed] table {sampled-snr-drm-best-120-816.dat};
		\addlegendentry{$\drrand$ (best)};
		\addplot[black, thick, dotted, very thick] table {sampled-snr-drm-worst-120-816.dat};
		\addlegendentry{$\drrand$ (worst)};
		\addplot[black, dashdotted, thick] table {sampled-snr-drm-dft-120-816.dat};
		\addlegendentry{$\drdft$};
		\end{axis}
		\end{tikzpicture}}
		\caption{Comparison of the three choices of \DRM $\dr$ for two different codes in two different scenarios. We include two realizations of the random choice, to see how the performance differs between realizations. In this scenario we consider $|r|=0.9$ and a cell edge \SNR of $-5$ \si{\dB}. The terminals are uniformly distributed (in angle) on the cell edge. The \DFT choice is poor, while the other choices perform similarly. The randomness stems from the user positions as well as the small-scale fading. a) $M=24$ \BS antennas, using code \codetwo; b) $M=120$ \BS antennas, using code \codeeight.}\label{fig:snr:drm}
\end{figure}
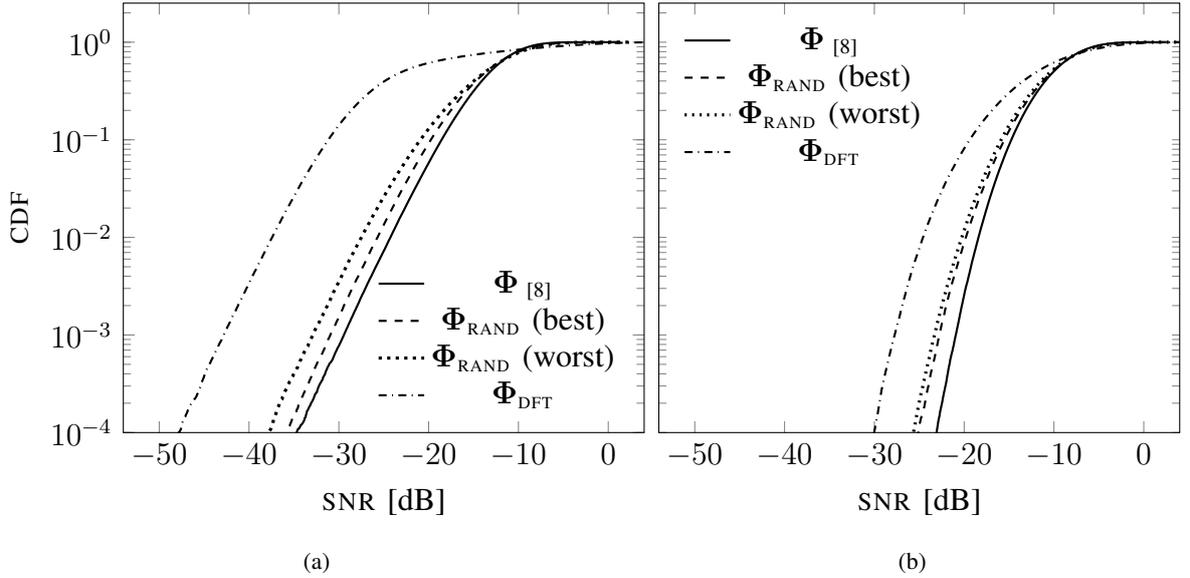

To see the effects of the \DRM, consider a correlation coefficient $|r|=0.9$ for two scenarios: one where the \BS has $M=24$ antennas and uses code \codetwo, and one where the \BS has $M=120$ antennas and uses code \codeeight. \Figref{fig:snr:drm} shows the cumulative distribution function (\CDF) of the \SNR \Eqref{eq:snr:ostbc} for uniformly distributed users on the cell edge when the \BS is using different \DRMs. To see the variation in performance of $\drrand$, which is random by definition, \Figref{fig:snr:drm} shows the best and the worst out of $10$ realizations.

The difference in performance is solely due to the different \DRM and how well these ``match'' the covariance matrix. The randomness is due to user positions and the small-scale fading. We see that $\drdft$ performs poorly here, giving some users very good performance, and some very poor. In general, \systeminfo should be available to as many users as possible, so preferably the curves should be vertical (and far to the right). That is, a spatially selective \DRM, with a large (approximative) null space performs poorly when the terminals are uniformly distributed.

Interestingly, the random choice performs at a similar level as the optimized \DRM in terms of symbol \SNRs. That being said, $\drrand$ does not satisfy, for example, the constraint necessary to ensure equal power through all antennas as $\dromni$ does. In addition $\dromni$ performs slightly better than $\drrand$ for larger codes, as seen in \Figref{fig:snr:drm:eight}. Nevertheless picking a random \DRM works relatively well. For larger codes, the performance of $\drdft$ improves, but is always considerably worse than both $\drrand$ and $\dromni$. This is due to the mismatch between $\drdft$ and the covariance matrix $\Cg$. If the covariance matrix has a different structure or if users are distributed differently, the \DFT choice might very well perform similar to or better than the other choices.

Looking at figures similar to \Figref{fig:snr:drm} for different scenarios (different $ M$, $|r|$, and codes, not included here) more general conclusions can be drawn. $\dromni$ is a ``one size fits all'' \DRM. It performs well for many choices of channel covariance matrices, codes and number of transmitting antennas. However, this does not mean that it is optimal in the sense of offering coverage to the largest number of terminals for any channel.

\section{Performance Metric}

To evaluate the performance of different codes in various settings, we consider outage rates instead of ergodic measures on capacity because of the limited number of diversity branches. It was shown in \cite{Yang14} that 
\[ \rstar = \outagecap + \ordo\left(\dfrac{\log(n)}{n}\right),  \]
where $\outagecap$ denotes the outage capacity and $\rstar$ denotes the maximal achievable rate for block length $n$ and outage probability $\epsilon$. That is, the outage capacity is a good approximation to $\rstar$ if $n$ is large enough.

An additive white Gaussian noise (\AWGN) channel with an \SNR of $x$ can reliably support a maximum rate of $\log_2(1+x)$ \cite[Section 5.4.1]{Tse05}. This means, conditioned on the channel estimate $\hhat$ and assuming worst-case noise (Gaussian), the effective channel in \Eqref{eq:detect:complex} can support a maximum rate of 
\begin{equation}\label{eq:maximum:supported rate}
\dfrac{\ns}{\td}\log_2\left(1 + \snr^{\ostbc}\right)\ \bpcu.
\end{equation}
Outage occurs if the used rate $R$ is larger than \Eqref{eq:maximum:supported rate}, i.e., if 
\[\rate > \dfrac{\ns}{\td}\log_2(1+\snr^{\ostbc}).\] 
The received symbol \SNR at the terminal depends on the realization of the channel estimate which in turn depends on the true channel. We assume independent channel realizations in each coherence interval and let $\snr^{\ostbc}_{l}$ denote the \SNR experienced at the terminal in coherence interval $l$ when an \ostbc is used at the \BS. Assuming coding over $L$ different coherence intervals, the average supported rate is 
\[ \frac{1}{L}\sum\limits_{l=1}^{L} \dfrac{\ns}{\td}\log_2\left(1 + \snr^{\ostbc}_{l}\right)\ \bpcu.\]
The probability of outage when using a rate $\rate$ is then 
\[ \outageprob^{\ostbc}(R) \triangleq \Pr\left(\rate > \frac{1}{L}\sum\limits_{l=1}^{L} \dfrac{\ns}{\td}\log_2\left(1 + \snr^{\ostbc}_{l}\right)\right). \]
For a given $\epsilon$, the outage capacity is defined as
\[ \outagecap^{\ostbc} \triangleq \sup \{R: \outageprob^{\ostbc}(R) < \epsilon \}. \]
In order to take training into account, we define the outage rate as
\begin{equation}\label{eq:outage:rate:ostbc}
\Reps^{\ostbc} \triangleq \left(\dfrac{\tc-\tp}{\tc}\right)\outagecap^{\ostbc} \ \bpcu,
\end{equation}
where we have scaled the outage capacity by the fraction of the coherence interval used for transmitting data. 

Completely analogous to \Eqref{eq:outage:rate:ostbc} we can define outage rates for general transmission and for transmission with a square \ostbc, using \Eqref{eq:snr:general} and \Eqref{eq:snr:square}, respectively. We let $\snr_l^{\text{general}}$ and $\snr_l^{\text{square}}$ be the \SNR experienced by the terminal in the $l$:th coherence interval in the two cases. Performing identical calculations as above gives the corresponding outage rates
\begin{equation}\label{eq:outage:rate:general}
\Reps^\text{general} \triangleq \left(\dfrac{\tc-\tp}{\tc}\right)\outagecap^\text{general} \ \bpcu,
\end{equation}
and
\begin{equation}\label{eq:outage:rate:square}
\Reps^\text{square} \triangleq \left(\dfrac{\tc-\tp}{\tc}\right)\outagecap^\text{square} \ \bpcu.
\end{equation}
We expect that $\Reps^\text{general} \geq \Reps^{\ostbc}$, which we will quantify numerically in \Secref{sec:simulations}.

One thing to note about the rate in \Eqref{eq:maximum:supported rate} is that the reciprocal of the code rate $\dfrac{\ns}{\td}$ is found in the numerator of $\snr^{\ostbc}$. For low \SNR this means that the rate in \Eqref{eq:maximum:supported rate} is almost independent of the code rate, since $\log(1+x)\approx x$ if $x\ll 1$.

\section{Simulations}\label{sec:simulations}

We consider the \ostbcs listed in \Tableref{tab:codes} and compare the outage rates of these, as defined in \Eqref{eq:outage:rate:ostbc}, in different scenarios. We will see how the performance varies depending on the number of \BS antennas, $M$, and the correlation coefficient $|r|$. In the end, we will also compare the results of the single-cell case to that of a multi-cell case.

Throughout the simulations, the outage probability $\epsilon=0.01$ is fixed. The terminals are distributed uniformly in a disk with radius $1$ in the single-cell case and in a regular hexagon with circumradius $1$ in the multi-cell case. Both in the single and multi-cell case, a small disk with radius $0.035$ around the \BS is excluded. Large-scale fading consists of distance-dependent path loss with exponent $3.8$ and the cell edge \SNR is set to $-5$ \si{\dB}. The coherence interval consists of $\tc = 256$ symbols.\footnote{The specific number was chosen to be a power of two, to simplify some of the simulations. It is still in the same order of magnitude as the smallest scheduling unit in \LTE ($168$) and the coherence interval for a coherence time of $1$ ms and a coherence bandwidth of a few hundred kHz.} We only consider \DRM $\dromni$, as this performs well in all tested scenarios.

Initially, we will only consider transmission over one coherence interval; hence no time/frequency diversity is available. In Sections \ref{sec:simulations:with:diversity}, \ref{sec:simulations:fixed:message} and \ref{sec:simulations:fixed:uses}, the \BS is allowed to code over several coherence intervals. Results from the multi-cell scenario is presented in \Secref{sec:simulations:multi:cell}.

\subsection{Pilot Energy Optimization}\label{sec:optimization}

To facilitate fair comparisons, all transmission strategies---no matter what code or \DRM---will have the same energy budget (the amount of energy spend in one coherence interval). We consider a heuristic way of optimizing the pilot energy, $\tp\pp$, by maximizing the outage rate of a simplified scenario, with the same parameters. We only optimize over $\pp$ since \cite[Theorem 1]{Cheng16} ensures that the outage rate is maximized when $\tp=\nt$.

To perform the heuristic optimization, the \BS assumes that a square \ostbc is used, the channel coefficients are i.i.d., and that the \SNR at the user only depends on the large-scale fading, which has a known distribution. Note that the optimization can be done regardless of the validity of these assumptions. Now, with these assumptions, the outage rate is given by~\Eqref{eq:outage:rate:square}. For an outage probability of $\epsilon$, the \BS considers the large-scale coefficient associated with the $\epsilon$ percentile, denoted $\beta_{\epsilon}$. That is, a fraction $1-\epsilon$ of the large-scale fading coefficients is larger than $\beta_{\epsilon}$ and a fraction $\epsilon$ is smaller than $\beta_{\epsilon}$. The \BS then considers the outage rate in \Eqref{eq:outage:rate:square} and calculates the value of $\pp$ such that this outage rate is maximized.  For our purpose, this heuristic method does not necessarily result in the optimal pilot energy because the resulting symbol \SNR \Eqref{eq:snr:ostbc} when using the codes in \Tableref{tab:codes} will not equal the symbol \SNR in \Eqref{eq:snr:square}. This method, however, does not require any \CSI at the \BS.

\begin{figure}[t]
	\centering
	\begin{tikzpicture}
	\begin{axis}[
	xlabel={received \SNR [dB]},
	ylabel={\CDF},
	ymode=log,
	table/y index={1},
	table/x index={0},
	table/col sep=comma,
	ymin=0.0001,
	xmax=50,
	legend style={at={(0.98,0.52)},anchor=south east},
	]
	\addplot[mark=none,solid, thick] table {sampled-snr-optimization-0-22.dat};
	\addlegendentry{$\pp=\pd$};

	\addplot[mark=none, thick, dashed] table {sampled-snr-optimization-1-22.dat};
	\addlegendentry{opt. $\pp$}
	\addplot[mark=none,solid,black,thick] table {sampled-snr-optimization-0-816.dat};
	\addplot[mark=none,dashed,black, thick] table {sampled-snr-optimization-1-816.dat};
	\end{axis}
	
	\draw \boundellipse{1.3,1.7}{0.4}{0.1};
	\node at (0.7,1.8) {\codetwo};
	
	\draw \boundellipse{2.1,1.0}{0.4}{0.1};
	\node at (2.7,1.1) {\codeeight};
	\end{tikzpicture}
	\caption{The received \SNR \Eqref{eq:snr:ostbc} with and without optimizing the pilot energy for code \codetwo and code \codeeight. We see that heuristically optimizing the pilot energy is beneficial.}
	\label{fig:opt:snr}
\end{figure}
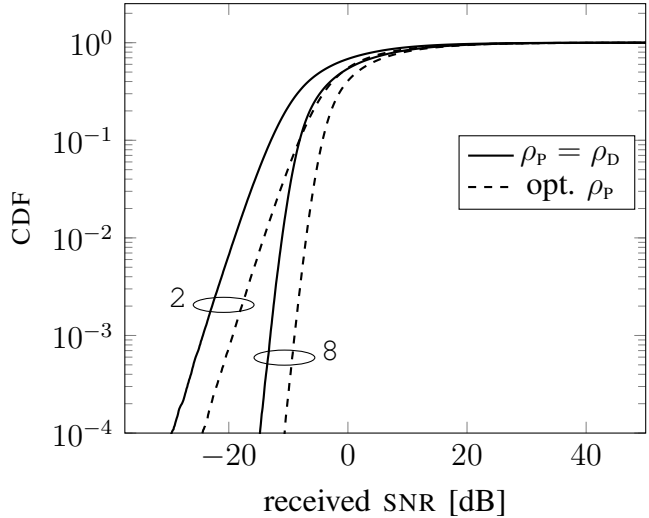

Now, let us see the effect of the optimization, by comparing the performance to the baseline: spending the minimum amount of symbols on pilots ($\tp=\nt$), while keeping the transmission power constant over the entire coherence interval ($\pd=\pp$). We consider the case of uncorrelated channels ($r=0$) here, but the same conclusions can be drawn when looking at correlated channels. \Figref{fig:opt:snr} shows the \CDF for the \SNRs with and without optimizing the pilot energy for two different codes. As seen, the baseline lags behind considerably for both codes, and the optimization proves useful.

In light of these results, all presented outage rates in the remainder of the paper have been optimized as presented in this section, which means that all codes use the minimum number of pilot symbols ($\tp=\nt$). As a consequence, the pilot symbols will be transmitted with considerably more power than the subsequent data symbols.

\subsection{Without Time/Frequency Diversity}

\begin{figure}[t]
	\subfloat[]{ %
	\label{fig:outage:rates:iid}
	\begin{tikzpicture}
	\begin{axis}[xlabel near ticks,
			     ylabel near ticks,
				table/x index={0},
				table/col sep=comma,
				ymode=log,
				ymax = 0.15,
				ymin = 0.009,
				xlabel={base station antennas, $M$},
       			ylabel={outage rate [bpcu]},
       			xtick={24,48,...,120},
       			legend style={at={(0.98,0.09)},anchor=south east}]
				\addplot[StyleCodeEight, samples=5, domain=24:120, forget plot] {0.126370};
				\addplot[StyleCodeTwelve, samples=5, domain=24:120, forget plot] {0.117364};
				\addplot[StyleCodeFour, samples=5, domain=24:120, forget plot] {0.105969};
				\addplot[StyleCodeTwo, samples=5, domain=24:120, forget plot] {0.0557161};
				\addplot[StyleCodeOne, samples=5, domain=24:120, forget plot] {0.0107173};
				\addlegendimage{empty legend};
				\addlegendentry{\hspace{-.6cm}\textbf{$\Reps^\text{general}$}};
				\addplot[mark=diamond*,only marks, samples=5, domain=24:120] {0.130516};
				\addlegendentry{$\nt=8$};
				\addplot[mark=pentagon*,only marks, samples=5, domain=24:120] {0.12291};
				\addlegendentry{$\nt=12$};
				\addplot[mark=square*,only marks, samples=5, domain=24:120] {0.105689};
				\addlegendentry{$\nt=4$};
				\addplot[mark=triangle*,only marks, samples=5, domain=24:120] {0.0550819};
				\addlegendentry{$\nt=2$};
				\addplot[mark=*,only marks, samples=5, domain=24:120] {0.0106272};
				\addlegendentry{$\nt=1$};
	\end{axis}
	\end{tikzpicture}}
	\subfloat[]{ %
	\label{fig:outage:rates:correlated}
	\begin{tikzpicture}
	\begin{axis}[xlabel near ticks,
				table/x index={0},
				table/col sep=comma,
				ymode=log,
				ymax = 0.15,
				ymin = 0.009,
				xlabel={base station antennas, $M$},
				ymajorticks=false,
			     xtick={24,48,...,120},
			     legend style={at={(0.98,0.10)},anchor=south east},
			     ] %
				\addlegendimage{empty legend};
							\addlegendentry{\hspace{-.6cm}\textbf{$\Reps^\text{\ostbc}$}};
							\addplot[StyleCodeEight] table[y index={3}] {outage-correlated.csv};
							\addlegendentry{\codeeight};
							\addplot[StyleCodeTwelve] table[y index={4}] {outage-correlated.csv};
							\addlegendentry{\codetwelve};
							\addplot[StyleCodeFour] table[y index={2}] {outage-correlated.csv};
							\addlegendentry{\codefour};
							\addplot[StyleCodeTwo] table[y index={1}] {outage-correlated.csv};
							\addlegendentry{\codetwo};
							\addplot[StyleCodeOne, samples=5, domain=24:120] {0.0107173};
							\addlegendentry{\codeone};
	\end{axis}
	\end{tikzpicture}}
	\caption{The outage rates for uncorrelated and correlated fading. In both scenarios, the spatial diversity pays off a great deal, but reaches a clear point of diminishing return for the larger codes. a) The two outage rates \Eqref{eq:outage:rate:ostbc} and \Eqref{eq:outage:rate:general} are very tight for most choices of $\nt$, but we see a slightly larger difference for $\nt=12$, as the markers do not overlap completely. b) A correlated channel with correlation coefficient $|r| = 0.9$ is considered. As the number of antennas grows, the correlation becomes negligible since $|r|^{M}$ decays quickly and the outage rate approaches that of the uncorrelated channel. The correlation strikes the larger codes harder when the number of \BS antennas is small.}\label{fig:outage:rates}
\end{figure}
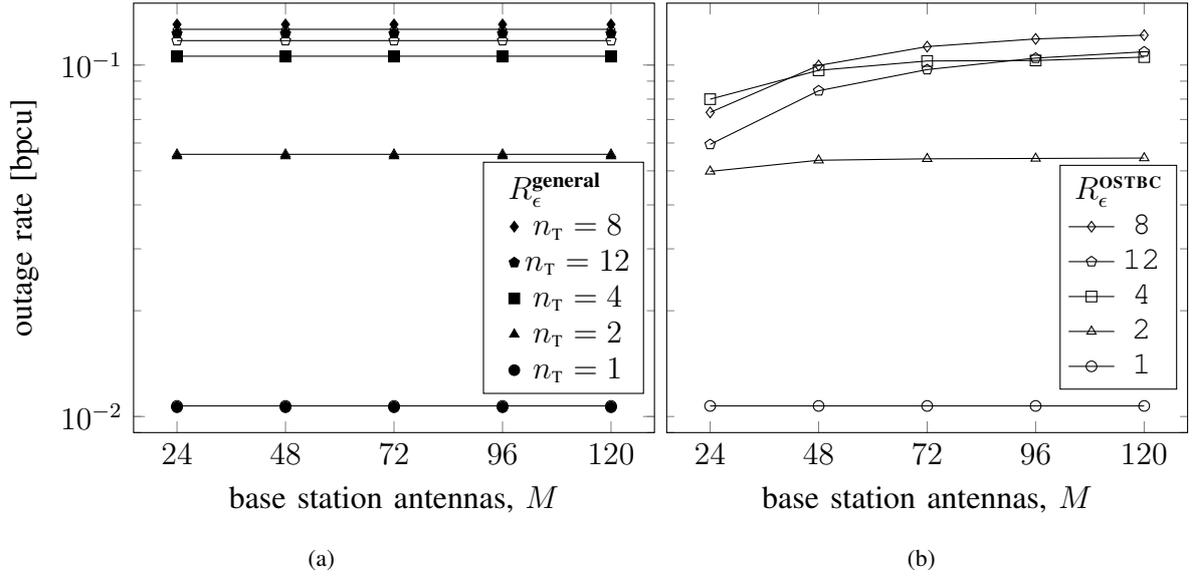

First, let us consider the case of i.i.d.~Rayleigh fading. The outage rates for the considered codes are shown in \Figref{fig:outage:rates:iid}. As indicated by theoretical results, the performance does not depend on the number of antennas (nor the chosen \DRM, as long as it is semi-unitary). In the same graph, shown with filled markers, are the achievable outage rates for \Eqref{eq:outage:rate:general}. We first note that the two bounds are tight, not only for codes \codeone and \codetwo as we mentioned in \Secref{sec:transmission:data}, but also for rectangular codes with code rate less than one, as seen by the overlapping markers. This is because the \SNR is low here, so the decrease in code rate is compensated by the increase in \SNR.

When time/frequency diversity is scarce, spatial diversity is extremely useful. Studying \Figref{fig:outage:rates:iid} more closely reveals that adding just a little spatial diversity can have a big impact, and the effect is more prominent the smaller outage probability, $\epsilon$, we require. Increasing the diversity order, going from $1$ to $2$ (effectively doubling the number diversity branches) gives a fivefold increase in outage rate. As we again double the diversity order, from $2$ to $4$, the rate is doubled. Doubling yet again, up to diversity order $8$, gives a moderate increase of about 10 percent. The diminishing return of diversity is most apparent when comparing the two larger codes. In \Figref{fig:outage:rates:iid}, the largest code does not give the highest rate. The reasons for this are twofold: First, the benefit of the extra spatial diversity is not big enough to counteract the effect of the increased pilot overhead. Second, the heuristic optimization works better for smaller codes (as the approximation of being square is more accurate). Around the point of $\nt = 10$, the effect of increasing the spatial diversity is overcome by the increase in pilot overhead, and thus larger codes are not useful. This is a consequence of the relatively short coherence interval, and the choice of outage probability $\epsilon$. Larger codes could still be useful in a scenario with longer coherence intervals or lower outage probability.

\Figref{fig:outage:rates:correlated} shows the outage rates for correlated channels with correlation coefficient $|r|=0.9$. When the channels are correlated, the outage rate decreases, as can be seen by comparing \Figref{fig:outage:rates:iid} and \Figref{fig:outage:rates:correlated}. This drop in performance is due to the \DRM not matching the channel covariance matrix when the channels are correlated, while any semi-unitary matrix matches the channel covariance matrix when the channels are uncorrelated. When the number of \BS antennas grows, the outage rate tends to that of the uncorrelated channel. This is because as the array grows, more antennas are further away from each other which decreases the correlation between the channels. Since $|r|^{M}$ decays quickly, only a moderate number of antennas is needed to mitigate even quite large correlation coefficients. The smaller codes struggle because of the lack of diversity, while the larger code gets punished by the symbols spent on pilots, as well as the optimization.

\subsection{With Time/Frequency Diversity}\label{sec:simulations:with:diversity}

Choosing the code giving the maximum rate, we see from \Figref{fig:outage:rates:iid} that the \BS can convey about $0.12$ \bpcu for the chosen scenario. Over one coherence interval this means about $30$ bits. If the \BS needs to convey more bits with the same outage probability, more resources have to be allocated.

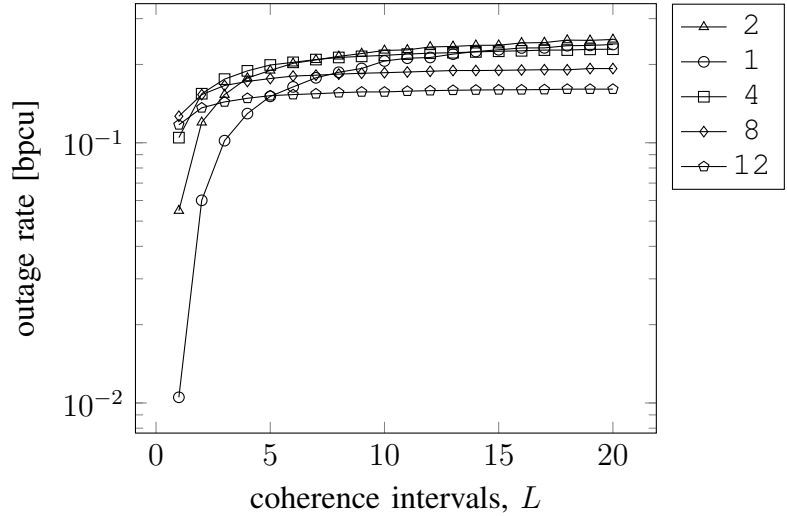
\begin{figure}[t]
	\centering
	\begin{tikzpicture}
		\begin{axis}[
			xlabel={coherence intervals, $L$},
			ylabel={outage rate [bpcu]},
			ymode=log,			
			legend pos=outer north east,
			table/x index={0},
			table/col sep=comma,
			]
			\addplot[StyleCodeTwo] table[y index={2}] {branches-bpcu.csv};
			\addlegendentry{\codetwo}
			\addplot[StyleCodeOne] table[y index={1}] {branches-bpcu.csv};
			\addlegendentry{\codeone}
			\addplot[StyleCodeFour] table[y index={3}] {branches-bpcu.csv};
			\addlegendentry{\codefour}
			\addplot[StyleCodeEight] table[y index={4}] {branches-bpcu.csv};
			\addlegendentry{\codeeight}
			\addplot[StyleCodeTwelve] table[y index={5}] {branches-bpcu.csv};
			\addlegendentry{\codetwelve}
		\end{axis}
	\end{tikzpicture}
	\caption{The outage rate \Eqref{eq:outage:rate:ostbc} when coding over several coherence intervals. The smaller codes performs poorly when only a few coherence intervals are allocated for transmission, because of the lack of diversity. As the number of allocated coherence intervals increases, the spatial diversity of the code matters less, and the codes with the highest code rate perform the best.}\label{fig:branches:bpcu}
\end{figure}

As we have seen previously, when the channel offers no time/frequency diversity, the larger codes tend to give a higher rate, since the spatial diversity from the code is so valuable. When the channel offers more time/frequency diversity, however, the spatial diversity from the code decreases in value. This is observed in \Figref{fig:branches:bpcu}, where the outage rate for each code is shown as a function the number of coherence intervals, $L$, the \BS codes over. Each coherence interval sees an independent channel realization, and hence the time/frequency diversity order is $L$.

In general, larger codes saturate faster, as they reach the point of diminishing returns quicker. They also saturate at a lower rate, because of the lower code rate, $\ns/\td$. Code \codeone gains a lot from the extra time/frequency diversity and quickly catches up to the other codes as the number of diversity branches increases. As $L$ tends to infinity, in which case ergodic capacity would be a relevant metric, performance is determined by the code rate, and hence, the smaller codes with higher code rate are superior. Note that the Alamouti code is better than \codeone for all values considered, as it offers more diversity at the same code rate.

\subsection{Fixed Message Length}\label{sec:simulations:fixed:message}

Ultimately, what code to choose depends on how much information the \BS needs to convey to the terminals. Consider a message of $\Nb$ bits. The \BS aims to reach $99$ percent (cf. $\epsilon=0.01$) of the terminals with this message. How many coherence intervals must be allocated to make this happen?

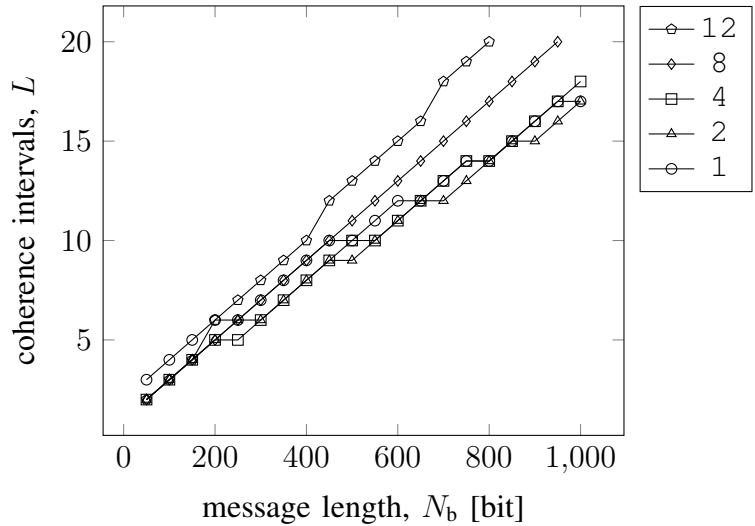
\begin{figure}[t]
	\centering
	\begin{tikzpicture}
		\begin{axis}[
			xlabel={message length, $\Nb$ [bit]},
			ylabel={coherence intervals, $L$},		
			legend pos=outer north east,
			table/x index={0},
			table/col sep=comma,
			]
			\addplot[StyleCodeTwelve] table[y index={5}, restrict y to domain=1:20] {resource-blocks-needed.csv};
			\addlegendentry{\codetwelve};
			\addplot[StyleCodeEight] table[y index={4}, restrict y to domain=1:20] {resource-blocks-needed.csv};
			\addlegendentry{\codeeight};
			\addplot[StyleCodeFour] table[y index={3}] {resource-blocks-needed.csv};
			\addlegendentry{\codefour};
			\addplot[StyleCodeTwo] table[y index={2}] {resource-blocks-needed.csv};
			\addlegendentry{\codetwo};
			\addplot[StyleCodeOne] table[y index={1}] {resource-blocks-needed.csv};
			\addlegendentry{\codeone};
		\end{axis}
	\end{tikzpicture}
	\caption{The minimum number of coherence intervals needed to convey a message of $\Nb$ bits with outage probability less than $\epsilon = 0.01$. For short messages, the larger codes provide sufficient rate, but as the message gets longer, the smaller code rate is too costly. For large messages, the base station needs to allocate more coherence intervals, providing time/frequency diversity, making the spatial diversity less useful. }\label{fig:branches:bits}
\end{figure}

We use the outage rates in \Figref{fig:branches:bpcu} and see how many bits can be conveyed using the different codes. Depending on the size of the message, $\Nb$, the \BS has to allocate different number of coherence intervals for each code. The minimum number of coherence intervals required for each code is shown in \Figref{fig:branches:bits}.

For many choices of message length $\Nb$, several codes might need the same number of coherence interval to convey the message, as seen by the overlapping curves in \Figref{fig:branches:bits}. In this case, we would choose the largest code, since the added diversity will make the received \SNR more reliable (slightly lower outage probability). The general trend is that larger codes are preferred for short messages, when few coherence intervals are needed, and smaller codes are preferred for long messages, as the many allocated coherence intervals provide enough diversity for the outage probability to be small. To take specific examples from \Figref{fig:branches:bits}, we see that code \codefour is preferred when $\Nb = 250$ and code \codetwo is preferred when $\Nb = 500$.

\begin{figure}[t]
	\centering
	\includegraphics{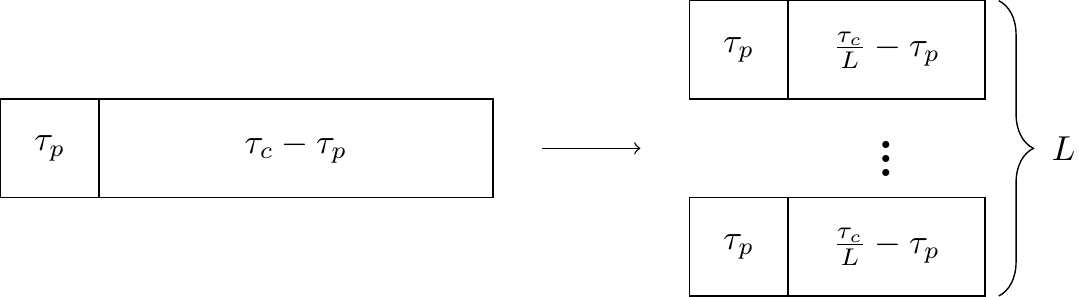}
	\caption{The $\tc$ channel uses does not necessarily have to be allocated in the same coherence interval. By spreading the \systeminfo over $L$ coherence intervals we get time/frequency diversity at a cost of increased pilot overhead.}\label{fig:coherence:split}
\end{figure}

\begin{figure}[t]
	\centering
	\begin{tikzpicture}
		\begin{axis}[
			xlabel={coherence intervals, $L$},
			ylabel={bits per $\tc$ channel uses},
			table/x index={0},
			table/col sep=comma,
			legend pos=outer north east,
			]
			\addplot[StyleCodeEight] table[y index={4}] {split-branches-bits.csv};
			\addlegendentry{\codeeight} 
			\addplot[StyleCodeTwelve] table[y index={5}] {split-branches-bits.csv}; 
			\addlegendentry{\codetwelve}
			\addplot[StyleCodeFour] table[y index={3}] {split-branches-bits.csv};
			\addlegendentry{\codefour}
			\addplot[StyleCodeTwo] table[y index={2}] {split-branches-bits.csv};
			\addlegendentry{\codetwo}
			\addplot[StyleCodeOne] table[y index={1}] {split-branches-bits.csv};
			\addlegendentry{\codeone}
		\end{axis}
	\end{tikzpicture}
	\caption{The total number of bits transferred over $\tc$ channel uses for the different codes in \Tableref{tab:codes} when transmission is split over several coherence intervals. Larger codes are punished quickly because of the relatively large pilot overhead, while smaller codes see an improvement due to their lack of diversity. However, approximately the same maximum is achieved regardless of what code is used.}\label{fig:branches:split:bits}
\end{figure}
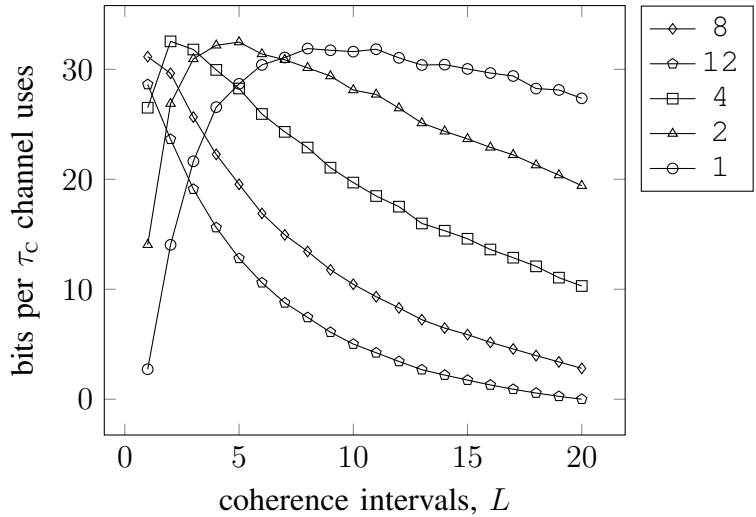

\subsection{Fixed Number of Channel Uses}\label{sec:simulations:fixed:uses}

We now allow for a coherence interval to carry both \systeminfo and other data. That is, the entire coherence interval does not necessarily have to be dedicated for \systeminfo. Although one coherence interval may carry both \systeminfo and other data, we do not multiplex spatially within one channel use as in \cite{Larsson16}. We analyze whether splitting up \systeminfo over several coherence intervals can improve performance. 

Consider having a total of $\tc=256$  channel uses dedicated to transmitting \systeminfo. If these channel uses are spread over several coherence intervals, we can code over several channel realizations, and hence the time/frequency diversity increases. On the other hand, we have to transmit downlink pilots in each of the coherence intervals, so fewer channel uses can actually be used for data. To be more precise: spreading the \systeminfo over $L$ coherence intervals will leave $\tc - L\tp$ channel uses for data, depicted in \Figref{fig:coherence:split}. This then yields a trade-off, once again, between diversity and pilot overhead, also mentioned in \cite{Yang12}. We stress that the minimum number of pilot symbols is used, i.e., $\tp = \nt$.

\Figref{fig:branches:split:bits} shows the total number of bits each code can transfer over $256$ channel uses, when transmission is spread over $L$ coherence intervals. The first thing to note is that all codes can, approximately, transfer the same amount of information, $31$ bits, over $\tc$ channel uses. This tells us that all codes perform similarly if the \BS is allowed to spread the \systeminfo over several coherence intervals. Second, the maximum for all codes occurs when the total number of diversity branches $L\nt$ is between $8$ and $10$. This means that, for this particular scenario, there is a tipping point at around $L\nt=9$ diversity branches: more branches require too much pilot overhead, fewer branches give too little diversity. This is why code \codetwelve performs worse than the others: the diversity is already saturated. The same phenomenon is observed for other scenarios as well, although the location of the tipping point differs. For a longer coherence interval or for a lower outage probability, the optimal number of diversity branches increases. As a consequence, the tipping point will move to the right.

\subsection{Multi-cell Setup}\label{sec:simulations:multi:cell}

We now consider a multi-cell setup with 19 cells: 18 interfering cells, and the home cell, in the center. We consider three different pilot-reuse factors and compare the outage rate when using different \ostbcs. Apart from now considering multiple cells, the setup is identical to that in \Figref{fig:outage:rates}, with the same correlation factor of $0.9$ and with $M=120$ \BS antennas. There are three important differences compared to the single-cell case, as mentioned in \Secref{sec:snr:multicell}: i) Contaminating cells that use the same pilots interfere with the channel estimation. This can be mitigated by increasing the pilot reuse. ii) The data transmitted by other cells increase interference in the symbol detection, and is independent of the pilot reuse. iii) An increased pilot reuse requires longer pilots and therefore increases the pilot overhead.

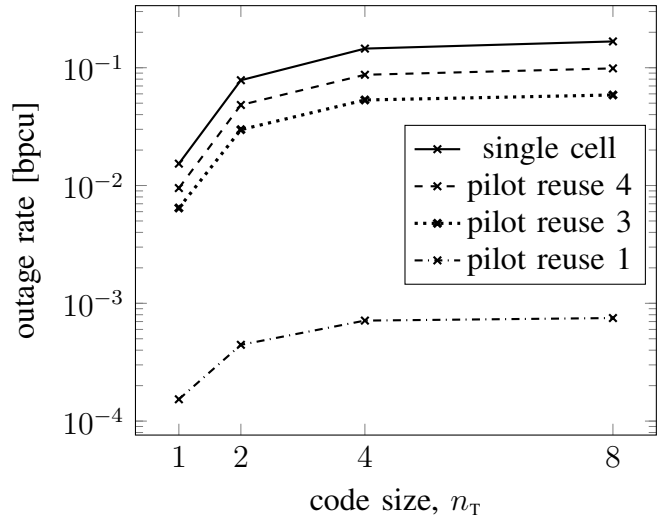
\begin{figure}[t]
\centering
	\begin{tikzpicture}
		\begin{axis}[
		xlabel={code size, $\nt$},
		ylabel={outage rate [bpcu]},
		table/col sep=comma,
		table/x index={0},
		ymode=log,
		mark options={solid},
		xtick={1,2,4,8},
		legend style={at={(0.98,0.35)}, anchor=south east},
		]
			\addplot[mark=x,solid, thick] table[y index={1}] {outage-single-hex.csv};
			\addlegendentry{single cell};
			\addplot[mark=x,dashed, thick] table[y index={3}] {outage-multicell-code.csv};
			\addlegendentry{pilot reuse 4};
			\addplot[mark=x,dotted, very thick] table[y index={2}] {outage-multicell-code.csv};
			\addlegendentry{pilot reuse 3};
			\addplot[mark=x, dashdotted, thick] table[y index={1}] {outage-multicell-code.csv};
			\addlegendentry{pilot reuse 1};
		\end{axis}
	\end{tikzpicture}
	\caption{Outage rates in a multi-cell systems with different pilot reuse for the four smallest codes in \Tableref{tab:codes}. The interference from cells using the same pilots is very strong in the case of pilot reuse 1, leading to a very low rate.}\label{fig:multicell}
\end{figure}

In \Figref{fig:multicell}, we see that the pilot reuse has a huge effect on the outage rates in a multi-cell system. When all cells use the same pilots, the outage rate is only a small fraction of what it is for the single-cell case. For pilot reuse 3 or 4, the outage rate is more similar to that of the single-cell. To make comparison fair here the shape of the single cell is hexagonal.

A secondary effect that also lowers the outage rates for the multi-cell setup is that the heuristic optimization in \Secref{sec:optimization} does not work as well as in the single-cell setup. This is because the effective \SNR experienced near the cell edge is much lower than what the heuristic method assumes (since it ignores all inter-cell interference). As a consequence, it is actually better to not optimize when using reuse 1 in our case.

\section{Conclusion}

Downlink transmission in massive \MIMO without \CSI at the base station, is necessary for conveying system information to the terminals in the cell. A massive \MIMO base station can outperform a single-antenna base station, with the same power constraint, in scenarios with and without correlated channels. Hence, conveying system information without \CSI is not a show-stopper for massive MIMO. As the number of diversity branches of the channel increases the benefit of the spatial diversity provided by the code decreases, making the larger codes primarily useful when time/frequency diversity is low. To convey short messages of a few hundred bits, less time-frequency resources are required and increased reliability can be provided if the base station uses codes which provide spatial diversity.
\bibliographystyle{IEEEtran} 
\bibliography{IEEEabrv,in-band-info-16.bib} 

\end{document}